\documentclass[aps,prc,superscriptaddress,twoside,twocolumn,nofootinbib,showpacs]{revtex4-1}
\usepackage{amsmath,amssymb}
\usepackage{mathtools}
\usepackage{bbold}
\usepackage[usenames, dvipsnames]{xcolor}
\usepackage{braket}
\usepackage{graphicx,bm}
\usepackage{slashed}
\usepackage{booktabs}
\usepackage{tensor}
\usepackage{multirow}
\usepackage{mwe}
\usepackage{changes}
\usepackage{enumerate}
\usepackage{stackrel}

\allowdisplaybreaks

\usepackage{epsfig}
\usepackage{epstopdf}
\usepackage{subfigure}
\usepackage{float}

\allowdisplaybreaks

\begin{document}


\title{Normalizing flows for microscopic many-body calculations: an application to the nuclear equation of state}

\author{Jack \surname{Brady}}
\email{jack.brady@tamu.edu}
\affiliation{Texas A\&M University, College Station, TX 77843, USA}

\author{Pengsheng Wen}
\email{pswen2019@tamu.edu}
\affiliation{Cyclotron Institute, Texas A\&M University, College Station, TX 77843, USA}
\affiliation{Department of Physics and Astronomy, Texas A\&M University, College Station, TX 77843, USA}

\author{Jeremy W. \surname{Holt}}
\email{holt@physics.tamu.edu}
\affiliation{Cyclotron Institute, Texas A\&M University, College Station, TX 77843, USA}
\affiliation{Department of Physics and Astronomy, Texas A\&M University, College Station, TX 77843, USA}

\date{\today}

\begin{abstract}
Normalizing flows are a class of machine learning models used to construct a complex distribution through a bijective mapping of a simple base distribution. We demonstrate that normalizing flows are particularly well suited as a Monte Carlo integration framework for quantum many-body calculations that require the repeated evaluation of high-dimensional integrals across smoothly varying integrands and integration regions. As an example, we consider the finite-temperature nuclear equation of state. An important advantage of normalizing flows is the ability to build highly expressive models of the target integrand, which we demonstrate enables precise evaluations of the nuclear free energy and its derivatives. Furthermore, we show that a normalizing flow model trained on one target integrand can be used to efficiently calculate related integrals when the temperature, density, or nuclear force is varied. This work will support future efforts to build microscopic equations of state for numerical simulations of supernovae and neutron star mergers that employ state-of-the-art nuclear forces and many-body methods.
\end{abstract}

\maketitle

{\it Introduction:}
The hot and dense matter equation of state (EOS) is of fundamental importance for interpreting observations of neutron stars, core-collapse supernovae, and neutron star mergers in terms of the underlying nuclear microphysics \cite{lattimer00,lattimer16}. Due to the complexity of computing the free energy and its derivatives (to obtain the pressure, entropy, chemical potentials, etc.) across the wide range of ambient conditions encountered during simulations of supernovae and neutron star mergers, most equations of state in wide use by the simulation community are based on simplified mean field models of the nuclear force \cite{lattimer91,shen91,shen11,steiner13}. Since mean field theory is grounded in effective interactions fitted to the bulk properties of medium-mass and heavy nuclei, one loses connection to fundamental nuclear two- and many-body forces and the ability to estimate systematic uncertainties \cite{drischler20prc}. In addition, certain thermodynamic properties that are important for understanding the evolution of core-collapse supernovae, such as the temperature-dependent nucleon effective mass \cite{yasin20,schneider19}, are quite different in microscopic and mean field models \cite{donati94}. For these reasons, there is strong motivation to develop more microscopic descriptions of the nuclear equation of state based on realistic nuclear forces in beyond-mean-field-theory quantum many-body calculations.

Microscopic calculations of the free energy $F(n,T,Y_p)$ as a function of density $n$, temperature $T$, and composition (e.g., the proton fraction $Y_p$) have in recent years been computed from realistic two- and three-body chiral effective field theory (EFT) nuclear forces \cite{wellenhofer15,carbone18,drischler21}. However, the inclusion of the most sophisticated three-body forces \cite{bernard08,bernard11} and important high-order many-body perturbation theory corrections (such as third-order particle-hole diagrams) \cite{holt17,drischler17} require the evaluation of technically challenging multi-dimensional integrations and therefore have not yet been achieved in finite-temperature calculations. Moreover, the tabulation of an astrophysical equation of state (the free energy and its first and second derivatives) involves the repeated evaluation of these integrals across more than 1,000,000 phase space points in order to ensure numerical stability of supernova and neutron star merger simulations~\cite{compose}. Microscopic EOS tabulations suitable for astrophysical simulations are therefore computationally demanding and only recently have been carried out \cite{togashi17,lu19} using the Argonne $v_{18}$ NN potential and the Urbana IX three-body force, supplemented by a liquid drop model for describing the low-density inhomogeneous phase of nuclear matter.

A potential solution to the numerical challenges outlined above is adaptive Monte Carlo methods based on importance sampling \cite{Hahn_2005}, which have recently been employed \cite{drischler17} to calculate high-order perturbation theory corrections to the cold dense matter equation of state that were previously intractable. In importance sampling an estimate for the integral
\begin{equation}
I[\psi] = \int_D \psi(\vec x) d\vec x
\label{int}
\end{equation}
is obtained as a Monte Carlo estimate under a proposal distribution $p(x)$ as
\begin{equation}
I \simeq \langle I \rangle_N = \frac{1}{N}\sum_{i=1}^N \frac{\psi(\vec x_i)}{p(\vec x_i)}.
\label{aint}
\end{equation}
The precision of this estimator, however, is dependent on how well $p(x)$ is able to match the normalized target $|\psi(\vec x)|/ \tilde I$, where $\tilde I \equiv I[|\psi|].$ In particular, if $p(x)$ matches $|\psi(\vec x)|/ \tilde I$ exactly, we obtain an ideal estimator~\cite{mcbook}. Consequently, when a precise estimate for the integral is required, as is the case when computing numerical derivatives in EOS tabulations, the proposal distribution must have sufficient expressive capacity to match the target. Popular adaptive importance sampling methods~\cite{Lepage1978ANA,Lepage2020AdaptiveMI}, however, often make restrictive assumptions on the integrand such as factorizability, thus limiting the precision of the estimator.

In the present work, we leverage normalizing flows~\cite{tabak10,tabak13,papamakarios19} as a means for constructing efficient importance sampling estimators for microscopic EOS tabulations. Normalizing flows have recently emerged as a highly expressive method for modeling complex proposal distributions~\cite{Rezende2015VariationalIW,Kingma2017ImprovedVI,Golinski2019AmortizedMC,mueller19,gao20,bothmann20,Wirnsberger2020TargetedFE,Kanwar2020EquivariantFS} using deep neural networks. We demonstrate that this expressivity allows for precise first- and second-order numerical derivatives of the free energy over a relatively coarse density and temperature grid. Furthermore, we show that a normalizing flow model trained on one target integrand transfers remarkably when the density, temperature, or nuclear force is varied, thus providing a compelling framework moving forward for including high-order many-body perturbation theory corrections for tabulated astrophysical equations of state and assessing associated uncertainties. Although the focus of the present work is on the nuclear matter equation of state, we note that normalizing flow based importance sampling could also be applied in condensed matter physics and related fields, where many-body perturbation theory has recently received renewed interest \cite{rossi17,macek20}.

As a concrete test case, we consider the second-order perturbation theory contribution to the grand canonical potential $\Omega$ of isospin-symmetric nuclear matter from an antisymmetrized two-body force $\bar V_{NN}$:
\begin{align}
& \Omega^{(2)} = -\frac{1}{8}\sum_{1234} \bar V_{NN}^{12;34} \bar V_{NN}^{34;12} \frac{f_1 f_2 \bar f_3 \bar f_4 - \bar f_1 \bar f_2 f_3 f_4}{\epsilon_3 + \epsilon_4 - \epsilon_1 - \epsilon_2},
\label{om2}
\end{align}
where $f_i = 1 / ( 1 + e^{(\epsilon_i-\mu)/T})$ is the Fermi-Dirac distribution function for particles with chemical potential $\mu$, $\bar f_i = 1 - f_i$, $\epsilon_i = k_i^2/(2M)$ is the free-particle spectrum, and the sums are taken over spin, isospin, and momentum. 
On the one hand, the contribution in Eq.\ \eqref{om2} is sufficiently complex to demonstrate the efficiency of normalizing flow based importance sampling, and on the other hand it is amenable to nearly exact evaluation using Gaussian quadrature for benchmarking our results. Since our focus is on the momentum-space integrations inherent in Eq.\ \eqref{om2}, we begin with a particularly simple model of the nuclear force
\begin{equation}
V(q) = \frac{g^2}{m_\phi^2+q^2}
\label{pot}
\end{equation}
associated with scalar-isoscalar boson exchange, where $q$ is the magnitude of the momentum transfer and we take $m_\phi=600$\,MeV and $g=1$. Performing the spin and isospin sums in Eq.\ \eqref{om2} and choosing $\vec k_3 = k_3 \hat i_z$, we obtain
\begin{align}
&\Omega^{(2)}(n,T) \nonumber \\
& = -\frac{M g^4}{64\pi^8} \int_0^{\infty} \!\!\! dk_1 \! \int_0^{\infty} \!\!\! dk_2 \! \int_0^{\infty} \!\!\! dk_3 \! \int_0^{\pi} \!\!\! d\theta_1 \! \int_0^{\pi} \!\!\! d\theta_2 \! \int_0^{2\pi} \!\!\!\!\! d\phi_1 \! \int_0^{2\pi} \!\!\!\!\! d\phi_2
\nonumber \\
& \frac{k_1^2 k_2^2 k_3^2 \sin \theta_1 \sin \theta_2 (f_1 f_2 \bar f_3 \bar f_4 - \bar f_1 \bar f_2 f_3 f_4)}{k_1^2 + k_2^2 - k_3^2 - k_4^2} \left [ \frac{4}{( m_\phi^2 + q_1^2)^2 }\right .\nonumber \\
& \left . -\frac{1}{( m_\phi^2 + q_1^2)( m_\phi^2 + q_2^2)} \right ] e^{-2(p/\Lambda)^6-2(p^\prime/\Lambda)^6},
\label{om2i}
\end{align}
where $\vec k_4 = \vec k_1 + \vec k_2 - \vec k_3$, $\vec q_1 = \vec k_1 - \vec k_3$, $\vec q_2 = \vec k_1 - \vec k_4$, $\vec p = \frac{1}{2}(\vec k_1 - \vec k_2)$, and ${\vec p}^{\, \prime} = \frac{1}{2}(\vec k_3 - \vec k_4)$. We have included the multiplicative function $g(\vec p,\vec p^{\, \prime}) = e^{-(\vec p/\Lambda)^6-(\vec p^{\, \prime}/\Lambda)^6}$ in the definition of the potential in Eq.\ \eqref{pot} as is common in the literature to regulate the unresolved high-momentum components of chiral nuclear forces \cite{machleidt11}. We choose $\Lambda = 450$\,MeV as the high-momentum cutoff scale. In practice, we have replaced the upper integration limits of $\infty$ with $k_{\rm max} = 2\Lambda$, which we have tested is sufficient to achieve converged results.

{\it Methods:}
A normalizing flow~\cite{tabak10,tabak13,papamakarios19} defines a complex distribution $p(\vec x)$ by applying a learnable, bijective mapping $\vec h$ to a simple base distribution $\pi(\vec u)$.
The probability density of a sample $\vec x \coloneqq \vec h(\vec u)$ under a flow can then be obtained analytically using the change of variables formula
\begin{equation}
p(\vec x) = \pi(\vec h^{-1}(\vec x)) \left | \det \left ( \frac{\partial \vec h^{-1}}{\partial \vec x} \right ) \right |.
\label{cov}
\end{equation}
In the present case, $\vec h$ should transform $\pi(\vec u)$ such that the resulting distribution $p(\vec x)$ matches our target distribution as closely as possible. This can be achieved by optimizing the parameters of $\vec h$ using gradient-based methods to minimize a suitable divergence metric between our model and target distributions. To optimize this objective in practice, however, certain conditions must be satisfied when choosing a parameterization for $\vec h$.

First, the transformation $\vec h$ must have sufficient expressive capacity to model the target distribution and at the same time have a Jacobian determinant that is tractable to compute. To satisfy these requirements, we implement $\vec h$ using a sequence of coupling transforms \cite{Dinh2015NICENI,Dinh2017DensityEU}. For a given coupling transform $\vec \phi$, an $n$-dimensional vector $\vec x$ is first partitioned into two parts: $\vec x  = (x_1,\dots,x_d,0,\dots 0) + (0,\dots,0,x_{d+1},\dots,x_n)$, which we refer to as the base input and updated input vectors, respectively. The $d$ coordinates of the base input vector are then passed through a neural network which outputs a set of parameters $\vec \theta$ that define an invertible transformation $g_{\theta_i}(x_i)$ dimension-wise on the $n-d$ updated input coordinates $x_i$. The coordinates of the base input vector $x_1,\dots,x_d$ are then passed unchanged through the coupling transform: $x_i \rightarrow x_i$, for $i\le d$. This results in a lower-triangular Jacobian matrix where the determinant can easily be computed as the product of the diagonals:
\begin{equation}
\det \left ( \frac{\partial \vec \phi}{\partial \vec x} \right ) = \prod_{i=d}^n \frac{\partial g_{\theta_i}}{\partial x_i}.
\end{equation}
Furthermore, $\vec h$ can be made highly expressive by composing a sequence of $k$ coupling transformations $\vec h$ := $\vec \phi_k$ $\circ$ $\dots$  $\circ$ $\vec \phi_1$ with different choices of base input and updated input coordinates such that all variables are allowed to interact.

The second condition $\vec h$ must satisfy is that its output range needs to respect the specified boundary conditions for a given integral.
For this purpose, we choose to implement each coupling transform using rational-quadratic spline flows~\cite{gregory82,mueller19,durkan19}, which define $g_{\theta_i}(x_i)$ piecewise on an interval $[A_i,B_i]$ by partitioning the interval into $K$ bins and defining the transformation in each bin as a monotonic rational-quadratic function. The rational-quadratic functions are parameterized by a set of $K+1$ knots $\{{(x^{(k)},y^{(k)})}\}_{k=0}^{K}$ which define the boundaries for the domain and range of each transformation, and a set of $K+1$ derivatives $\{{\delta^{(k)}}\}_{k=0}^{K}$ defined at each knot. The knots monotonically increase within the interval $[A_i,B_i]$ where ${(x_i^{(0)},y_i^{(0)})} = (A_i,A_i)$ and ${(x_i^{(K)},y_i^{(K)})} = (B_i,B_i)$ such that $g_{\theta_i}(x_i)$ is a mapping from $[A_i,B_i]$ to $[A_i,B_i]$. Thus, by setting $[A_i,B_i]$ to be the boundaries for each dimension of a given integral, we can restrict $\vec h$ to only be defined on the integration region.

We implement our flow using a composition of 6 rational-quadratic spline coupling transforms which is the minimum number required to account for correlations among all the variables in our seven dimensional integral in Eq.\ \eqref{om2i}~\cite{gao20}. For each transform, we use $K=16$ bins and implement each neural network using a residual network~\cite{He2016DeepRL} with two residual blocks and 32 hidden features. Our base distribution is chosen to be uniform over the integration region for each dimension, respectively. To train our flow, we minimize the Pearson $\chi^2$ divergence between our model distribution $p(\vec x)$ and target distribution $|\psi(\vec x)|/ \tilde I$. This divergence is estimated as an expectation under our model through importance sampling as
\begin{equation}
D_{\chi^2} \simeq \langle D_{\chi^2} \rangle_N = \frac{1}{N}\sum_{i=1}^N \frac{( \frac{|\psi(\vec x_i)|}{\tilde I} - p(\vec x_i))^2}{p(\vec x_i)} / p(\vec x_i),
\label{loss}
\end{equation}
where the normalizing constant $\tilde I$ is additionally estimated through sampling. At each training iteration, we minimize this expectation with respect to the parameters of our flow using gradient descent on batches of $N$ samples, where sampling from our flow amounts to first sampling from the base distribution $\pi(\vec u)$ and then passing these samples through $\vec h$ to obtain $\vec x$. The gradient descent optimization algorithm we employ is Adam~\cite{Kingma2015AdamAM}. All models were implemented using PyTorch~\cite{NEURIPS2019_9015}, and the open-source implementation for the spline transformations in~\cite{durkan19} was used for our coupling layers.

{\it Results:}
We start by training the flow on the target integral in Eq.\ \eqref{om2i} using batches of $N=5000$ samples drawn randomly at each iteration from our base distribution and passed through our model. We initially fix the density at $n=n_0$, where $n_0=0.16$\,fm$^{-3}$ is the saturation density of nuclear matter, and the temperature at $T= 25$\,MeV. The learning rate for the Adam optimizer was set to $10^{-3}$ until 200 iterations passed without an improvement in the $\chi^2$ loss function, at which point a cosine scheduler was initiated with maximum learning rate $10^{-3}$, minimum learning rate $10^{-4}$, and period of 200 iterations. For comparison, we have computed the integral in Eq.\ \eqref{om2i} using the adaptive Monte Carlo integrators Divonne, Suave, and VEGAS in the Cuba multidimensional integration library \cite{Hahn_2005}. Only VEGAS \cite{Lepage1978ANA,Lepage2020AdaptiveMI} was found to give a high-quality estimate of the integral as well as a reliable associated uncertainty, and therefore it will be the standard benchmark used throughout this work. Suave yielded inaccurate integral estimates, while Divonne was found to underestimate its actual error (see discussion below).

\begin{figure}[t]
     \centering
     \includegraphics[scale=0.5]{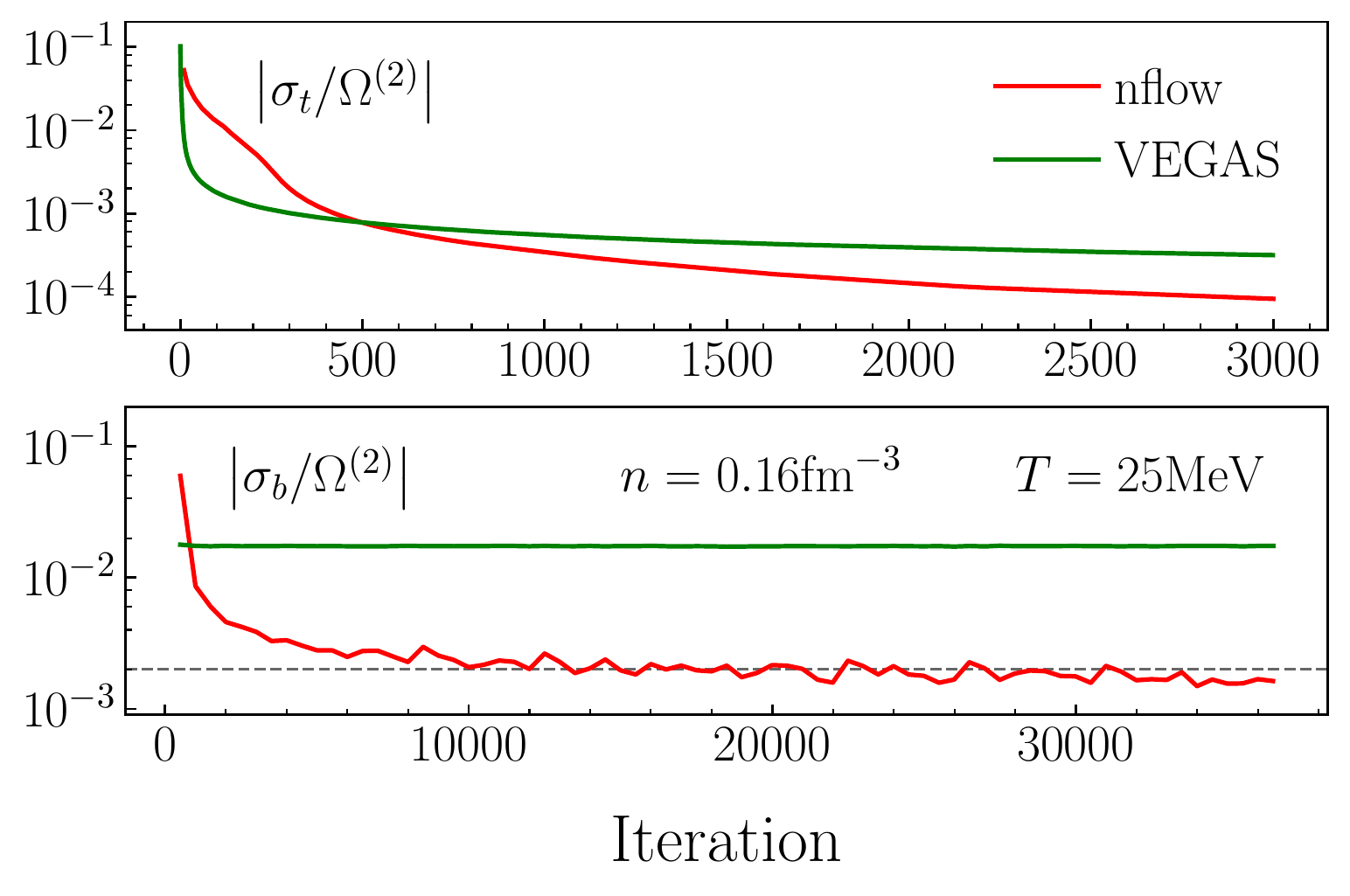}
     \caption{(top) Total relative uncertainty $|\sigma_{t}/\Omega^{(2)}|$ as a function of iteration at the beginning of training for the evaluation of $\Omega^{(2)}(n_0,T=25\,{\rm MeV})$ in Eq.\ \eqref{om2i} using the normalizing flow (red) and VEGAS (green) Monte Carlo integration algorithms.
     (bottom) Same as top panel but for the batch relative uncertainty $|\sigma_{b}/\Omega^{(2)}|$ as a function of iteration over the full training time. The number of batch samples per iteration is 5000. The gray dashed line is added to help guide the eye.
     }
	  \label{fig:it}
\end{figure}

In the top panel of Fig.\ \ref{fig:it} we compare the VEGAS (green) and normalizing flow (red) total relative uncertainty
\begin{equation}
    \frac{\sigma_t}{\Omega^{(2)}} = \frac{1/\sqrt{\sum_i \frac{1}{\sigma_i^2}}}{\sum_i \frac{\Omega^{(2)}_i}{\sigma_i^2}/{\sum_i \frac{1}{\sigma_i^2} }}
    \label{eq:relu}
\end{equation}
over the first 3000 iterations, where in Eq.\ \eqref{eq:relu} $i$ is the training iteration, $\sigma_i$ is the batch standard error, and $\Omega_i^{(2)}$ is the batch mean. We observe that VEGAS outperforms the normalizing flow early in the training, but the greater expressive capacity of the normalizing flow leads to a smaller total relative uncertainty past 500 iterations. Moreover, in stark contrast to VEGAS, the normalizing flow continues to learn over many training iterations, as shown in the bottom panel of Fig.\ \ref{fig:it}, where we plot the batch relative uncertainty $|\sigma_b/\Omega^{(2)}|$, which reflects the performance of each model at a particular training iteration with 5000 samples. We observe that the VEGAS batch uncertainty saturates after less than one hundred iterations, while the normalizing flow batch uncertainty continues to decrease throughout training. We stopped training the normalizing flow when the total relative uncertainty reached $10^{-5}$ (after 36,000 iterations) with associated batch relative uncertainty of $\sim 1.5 \times 10^{-3}$, which is an order of magnitude improvement over VEGAS.

\begin{figure}[t]
     \centering
     \includegraphics[scale=0.48]{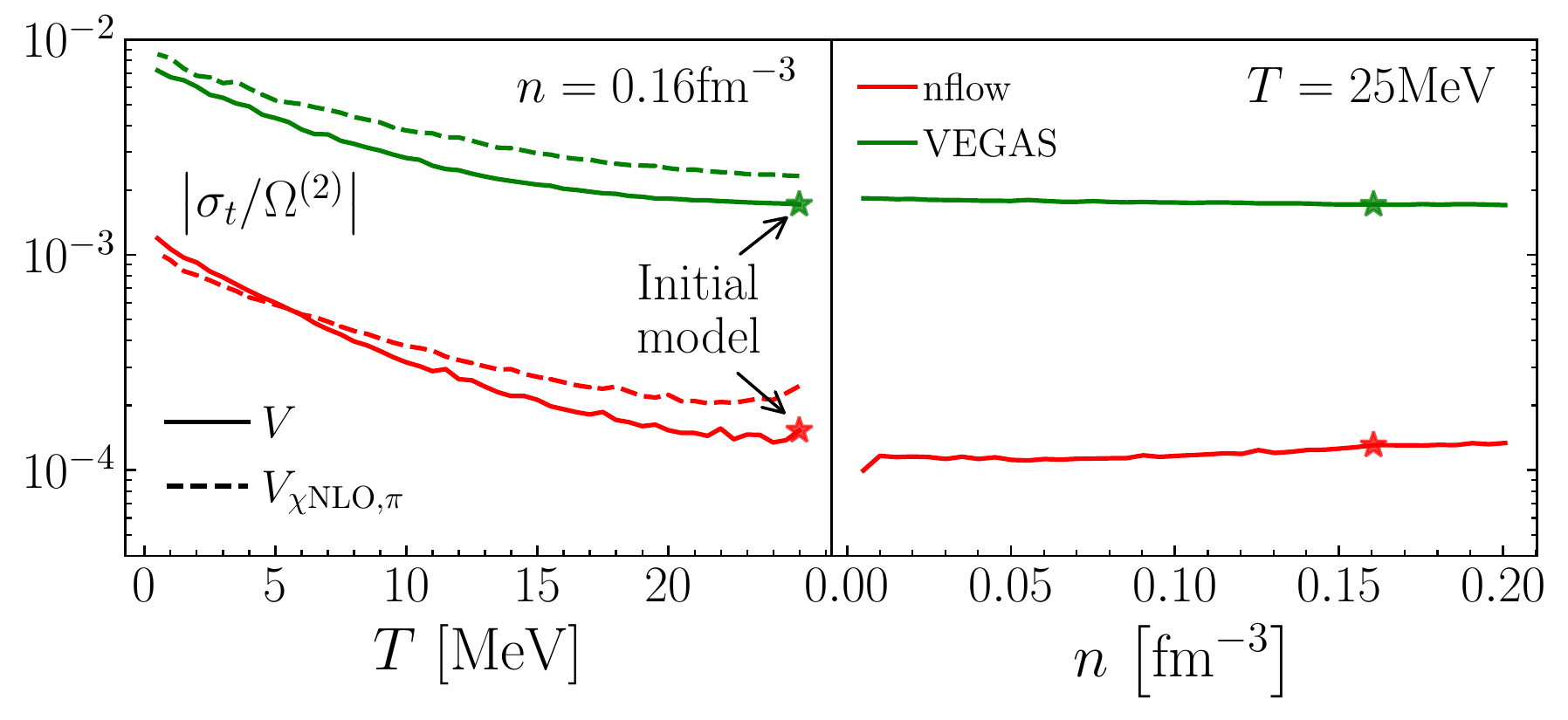}
     \caption{Total relative uncertainty for the evaluation of $\Omega^{(2)}(n,T)$ in Eq.\ \eqref{om2i} using the VEGAS (green solid line) and normalizing flow (red solid line) Monte Carlo integration algorithms. Both models were initially trained at $n=n_0$ and $T=25$\,MeV (denoted by the star) for 36,000 iterations and then transferred to nearby phase space points using only 100 additional training iterations (500,000 samples). Also shown are the total relative uncertainties (dashed lines) for both models trained for 100 iterations when the potential in Eq.\ \eqref{pot} is replaced by the pion-exchange terms up to next-to-leading-order in the chiral expansion.}
	  \label{fig:reluT}
\end{figure}

When precise integral estimates are required (as is the case for computing numerical derivatives of the free energy), it is clear that normalizing flows are able to outperform VEGAS, with the caveat that a high sample complexity is required to reach this low precision. We now demonstrate that this initial high sample complexity is a one-time cost, and that normalizing flow models transfer exceptionally well when either the density, temperature, or even the nuclear potential in Eq.\ \eqref{om2i} is varied. 
In the left and right panels of Fig.\ \ref{fig:reluT} we show the temperature and density dependence, respectively, of the total relative uncertainty $|\sigma_t/\Omega^{(2)}|$ in the evaluation of Eq.\ \eqref{om2i}. Both VEGAS (green) and the normalizing flow (red) were first trained at the phase space point indicated by the star and then transferred sequentially to different densities and temperatures using just 100 additional training steps (500,000 samples). At the starting point ($n=n_0, T=25\,{\rm MeV}$), the normalizing flow begins with more than an order of magnitude better uncertainty estimate compared to VEGAS. We can see that this improvement in the precision (relative to VEGAS) persists as both the density and temperature are varied. We note that the increase in total relative uncertainty as temperature decreases is not unexpected, since the sharpening of the Fermi distribution functions becomes difficult to model within any adaptive Monte Carlo method, as evidenced by the similar behavior demonstrated by VEGAS. 

We also show in Fig.\ \ref{fig:reluT} the ability for each model to adapt to nontrivial changes in the choice of nuclear potential. In particular, the dashed lines in the left panel of Fig.\ \ref{fig:reluT} denote the total relative uncertainty in the evaluation of Eq.\ \eqref{om2i} after replacing the simple potential in Eq.\ \eqref{pot} with the sum of the leading-order (LO) and next-to-leading-order (NLO) pion-exchange contributions in realistic chiral effective field theory nuclear forces \cite{epelbaum09rmp,machleidt11}. In obtaining the integral estimates for $V_{\chi {\rm NLO},\pi}$, we included one extra training run (100 iterations) at the phase space point ($n=n_0, T=25\,{\rm MeV}$) to reorient the normalizing flow model before using the standard 100 iterations to train and evaluate at all phase space points. We observe that the normalizing flow is able to efficiently transfer even when a highly nontrivial change to the nuclear potential is introduced.

\begin{figure}[t]
     \centering
     \includegraphics[scale=0.51]{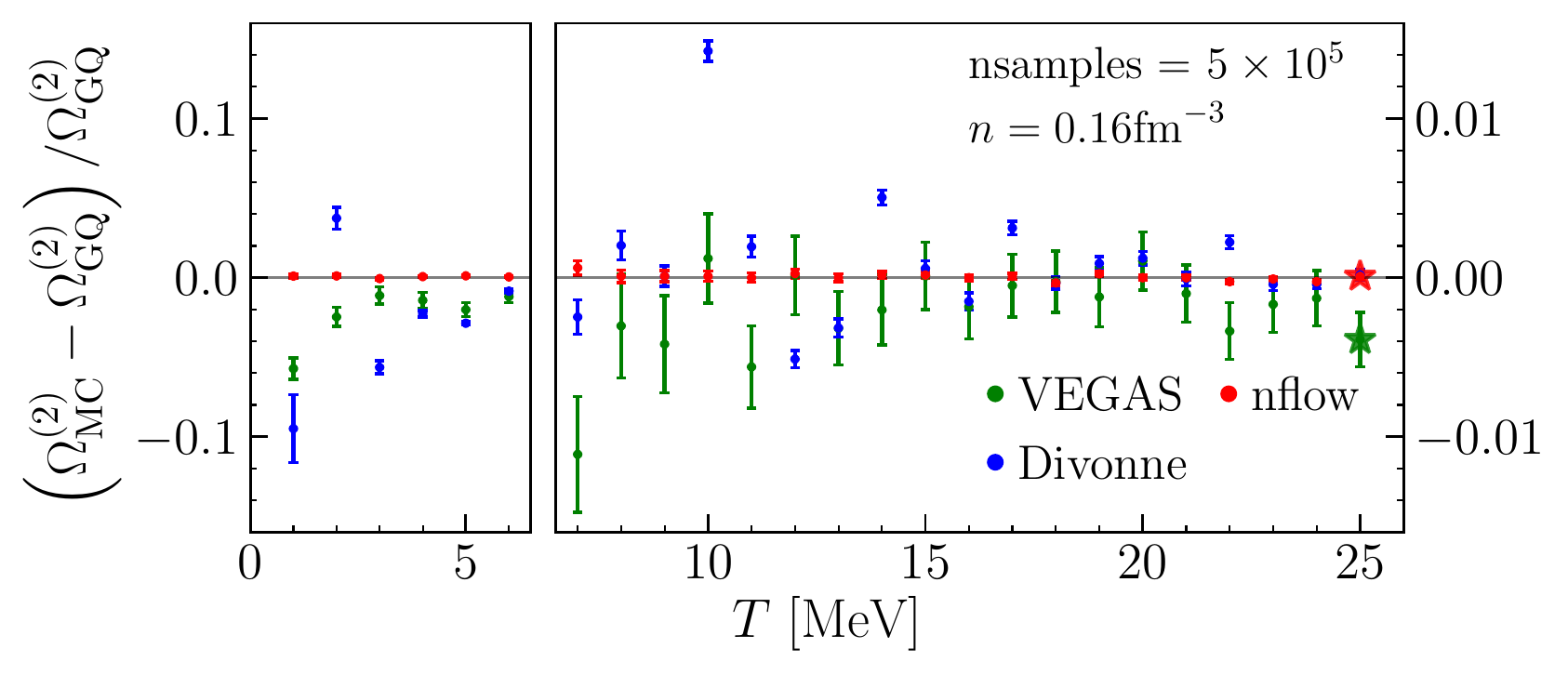}
     \caption{Relative error for the evaluation of $\Omega_{\rm MC}^{(2)}(n,T)$ in Eq.\ \eqref{om2i} using the VEGAS (green), Divonne (blue), and normalizing flow (red) Monte Carlo integration algorithms compared to the exact evaluation $\Omega_{\rm GQ}^{(2)}(n,T)$ using Gaussian quadrature. The VEGAS and normalizing flow models were initially trained at $n=n_0$ and $T=25$\,MeV and then transferred to lower temperatures using 100 additional training iterations (500,000 samples). }
	  \label{fig:relun}
\end{figure}

In Fig.\ \ref{fig:relun} we show the true Monte Carlo integration errors for Divonne, VEGAS, and normalizing flows by comparing to exact results obtained from Gaussian quadrature (GQ). In general, we observe that the integral estimates for $\Omega^{(2)}_{\rm MC}$ from the normalizing flow and VEGAS are within one or two standard deviations of the exact result $\Omega^{(2)}_{\rm GQ}$, while Divonne significantly underestimates its actual error.

\begin{figure}[t]
     \centering
     \includegraphics[scale=0.53]{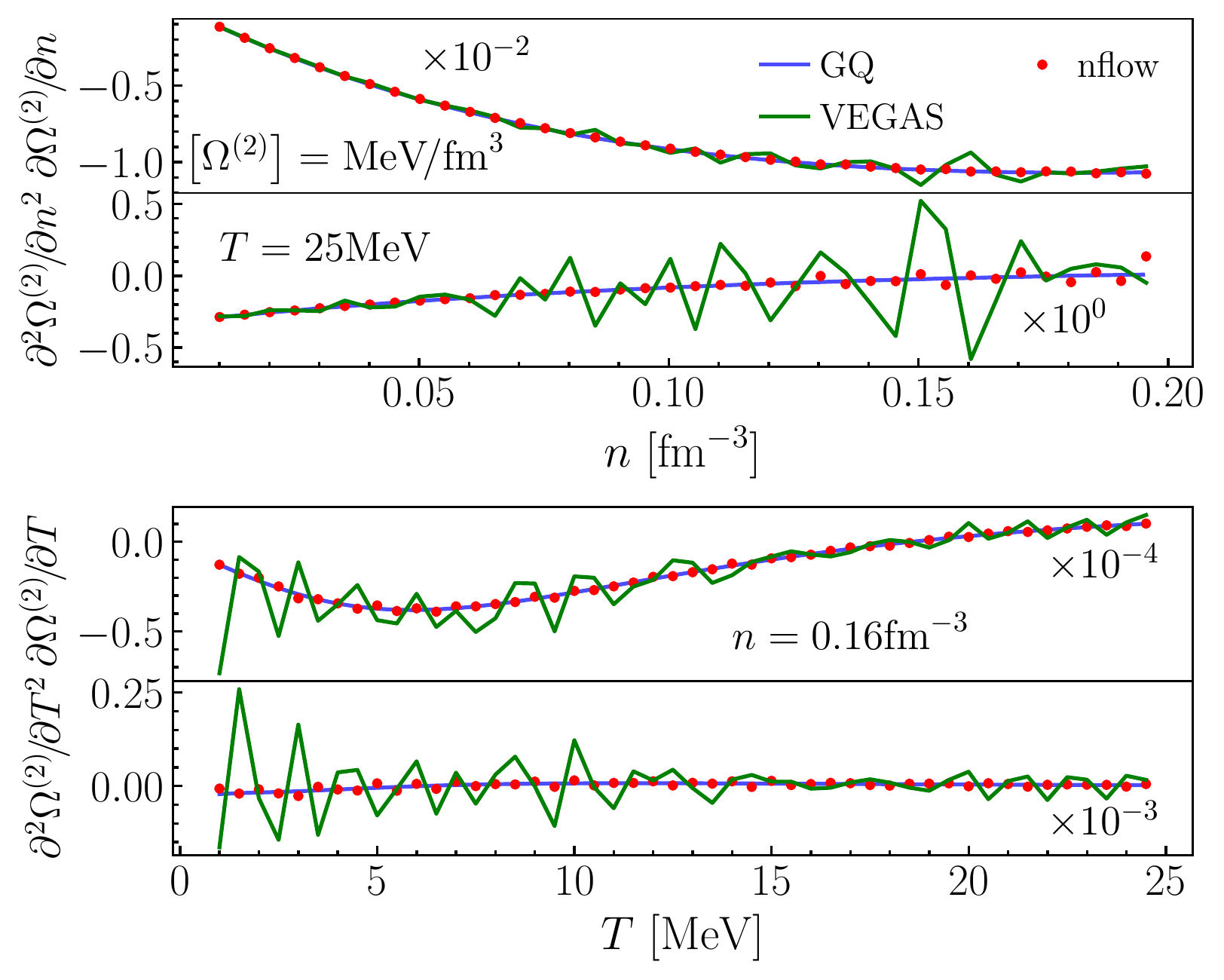}
     \caption{$1^{\rm st}$ and $2^{\rm nd}$-order derivatives of $\Omega^{(2)}$ with respect to the density $n$ and temperature $T$ from VEGAS (green line), normalizing flows (red dots), and exact Gaussian quadrature (blue line). The numerical derivatives are calculated by the central finite difference method of order 2 generated from the data shown in Fig.\ \ref{fig:reluT}.}
	  \label{fig:ddd}
\end{figure}

We now estimate how the uncertainties in the Monte Carlo estimates for the grand canonical potential from the VEGAS and normalizing flow models propagate to the calculation of numerical derivatives. In the top and lower panels of Fig.\ \ref{fig:ddd} we show the $1^{\rm st}$ and $2^{\rm nd}$-order derivatives of $\Omega^{(2)}$ with respect to the density and temperature, respectively. The central finite difference method of order 2 is applied to calculate the numerical derivatives from the VEGAS (green) and normalizing flow (red) datasets generated in Fig.\ \ref{fig:reluT} as well as the exact results obtained through Gaussian quadrature (blue). We see that the improved numerical precision from the normalizing flow leads to significantly better estimates of free energy derivatives. In particular, we can see that in the low-temperature region the derivatives from VEGAS fluctuate strongly about the true value, while the results from the normalizing flow are stable and match the exact values well even for the $2^{\rm nd}$-order derivatives.

{\it Outlook:}
In the present work, we have performed proof-of-principle calculations demonstrating the potential of normalizing flow based importance sampling in the context of nuclear many-body perturbation theory. In particular, we have shown that normalizing flows are able to learn models of the target integrand which allow for precise integral estimates and which can be transferred to related integrals with minimal additional computational cost. Ultimately, this leads to speedup factors on the order of 100 compared to VEGAS when 
precise integral evaluations must be repeated across a multi-dimensional phase space, such as in calculations of the free energy and its numerical derivatives in astrophysical equation of state tables. Numerous extensions and further applications are envisioned. One important application is the nucleon single-particle energy, a quantity that varies over the four-dimensional parameter space $\{n, T, Y_p, q\}$, where $q$ is the nucleon momentum. First and second-order derivatives of this quantity are needed when computing e.g., the nucleon effective mass. Another application is to nuclear matter response functions, which vary over the five-dimensional parameter space $\{n, T, Y_p, q, \omega\}$, where $q$ and $\omega$ represent the momentum and energy transfer to the medium. In all of these cases, normalizing flows may allow for the inclusion of perturbation theory contributions that at present are too computationally demanding to map over the full phase space needed in astrophysical applications.

\begin{acknowledgments}
J.B. would like to thank Prafulla Choubey for helpful collaboration in the early stages of this project. Work supported by the National Science Foundation under Grant No.\ PHY1652199 and by the U.S.\ Department of Energy National Nuclear Security Administration under Grant No.\ DE-NA0003841. Portions of this research were conducted with the advanced computing resources provided by Texas A\&M High Performance Research Computing.
\end{acknowledgments}

\bibliographystyle{apsrev4-1}
%

\end{document}